\documentclass[twocolumn,traditabstract,rnote]{aa}
\usepackage{graphicx}

\usepackage[T1]{fontenc}
\usepackage[a4paper]{geometry}
\makeatletter

\providecommand{\tabularnewline}{\\}

\usepackage{pdflscape}

\makeatother

\begin{document}

\title{The Canada-France High-$z$ Quasar Survey: 1.2\,mm Observations
}

\author{Alain Omont\inst{1,2}
\and Chris J. Willott\inst{3}
\and Alexandre Beelen\inst{4}
\and Jacqueline Bergeron\inst{1,2}
\and Gustavo Orellana\inst{5,4}
\and Philippe Delorme\inst{6}
}
%
\institute{UPMC Univ Paris 06, UMR7095, Institut d'Astrophysique de Paris, F-75014, Paris, France
\and  CNRS, UMR7095, Institut d'Astrophysique de Paris, F-75014, Paris, France 
\and Herzberg Institute of Astrophysics, National Research Council, 5071 West Saanich Rd, Victoria, BC V9E 2E7, Canada
\and  Univ Paris-Sud and CNRS, Institut d'Astrophysique Spatiale, UMR8617, Orsay, F-91405, France
\and Astronomy DepartmentUniversidad de Concepci\'on, Concepci\'on, Chile 
\and UJF-Grenoble 1/CNRS-INSU, Institut de Plan\'etologie et d'Astrophysique de Grenoble (IPAG) UMR 5274, 38041, Grenoble, France 
}




\abstract{
 We report 250\,GHz (1.2\,mm) observations of a sample of 20 QSOs 
 at redshifts $5.8<z<6.5$ from the the Canada-France High-z Quasar Survey 
(CFHQS), using the Max-Planck Millimeter Bolometer (MAMBO) array at the 
IRAM 30-metre telescope. A rms sensitivity $\lesssim$\,0.6\,mJy was 
achieved for 65\% of the sample, and $\lesssim$\,1.0\,mJy for 90\%. Only 
one QSO, CFHQS\,J142952+544717, was robustly detected with S$_{\rm 
250GHz}$\,=\,3.46$\pm$0.52\,mJy. This indicates that one of the most 
powerful known starbursts at $z\sim 6$  is associated with this radio 
loud QSO. On average, the other CFHQS QSOs, which have a mean 
optical magnitude fainter than previously studied SDSS samples of 
$z\sim 6$ QSOs, have a mean 1.2\,mm flux density
${\rm \langle S_{250GHz}\rangle}$\,=\,0.41$\pm$0.14\,mJy; such a 2.9-$\sigma$ average detection is hardly meaningful. It would correspond to
$\langle$L$_{\rm FIR}\rangle$\,$\approx$\,0.94$\pm$0.32\,10$^{12}$\,L$_{\sun}$, 
and an average star formation rate of a few 100's M$_{\sun}$/yr, depending on the IMF and a possible AGN contribution to $\langle$L$_{\rm FIR}\rangle$. This is 
consistent with previous findings of Wang et al. on the 
far-infrared emission of $z\sim 6$ QSOs and  extends them toward 
optically fainter sources.
}


\maketitle

\section{Introduction}
The highest-redshift QSOs known, at $z\gtrsim 6$, are fascinating objects, providing crucial clues about the growth of supermassive black holes (SMBH), their host galaxies and their environment, when the Universe was less than 1 Gyr old, toward the end of the reionization epoch. 
Black holes of several 10$^9$\,M$_{\sun}$ were already in place (Willott et al.\  2003, Kurk et al. 2007; Jiang et al.\ 2007; Mortlock et al.\ 2012). Such a rapid growth of the mass of early black holes puts extremely severe constraints on classical accretion Eddington limited by radiation pressure. It might point to the existence of a more efficient process for forming a massive black hole such as direct collapse  without fragmenting (Begelman et al.\ 2006; Volonteri 2012).

In this context, it is obviously critical to accumulate more information about SMBH growth and the parallel formation of the first massive galaxies in the most massive dark matter halos at the reionization epoch. An important point to elucidate is to determine the black hole to galaxy mass ratio at this epoch, and to see how it compares to the current value at $z\,=\,0$. A related piece of information is provided by the relationship between the galaxy growth by star formation and the QSO luminosity generated in black hole accretion. Both processes mainly result from gas transport to the center of the galaxy. They should contribute in building the eventual tight relation between the masses of the black hole and the galaxy, even if this relation is mainly ruled by QSO feedback onto the interstellar gas.

As the emission of young stars is mostly channeled into the far-infrared (FIR) by dust in such massive starburst galaxies, the FIR luminosity, L$_{\rm FIR}$, directly reflects the star formation rate (SFR). In the absence of multi-wavelength sampling of the FIR emission, one may assume a standard infrared spectral energy distribution (SED) (e.g.\ Wang et al.\ 2008), and thus  
directly derive L$_{\rm FIR}$ and SFR from measuring the continuum flux density at a single wavelength. At  $z\gtrsim 6$, a single continuum observation in the $\lambda$$\sim$1\,mm window efficiently probes the bulk of the rest-frame FIR emission and provides thus a simple estimate of L$_{\rm FIR}$ and SFR. In addition, such $\sim$1\,mm observations yield valuable estimates of the dust mass in the host galaxy which reflects a combination of the gas mass and the metallicity. 

After a series of such mm/submm studies of FIR properties of $z\,\sim2$-4 QSOs (e.g. Omont et al.\ 2001; 2003; Carilli et al.\ 2001; Priddey et al.\ 2003), this method was successfully applied by Wang et al. (2008; 2011a) to investigate the 
 FIR properties, 
inferred from 1.2\,mm observations, of 40 $z\sim 6$ QSOs mainly discovered from the Sloan Digital Sky Survey (SDSS) (Fan et al.\ 2006).  A significant fraction (30\%) of Wang et al.'s  
$z\sim 6$ sample is detected at 1.2\,mm, pointing to an excess of FIR emission dominated by a strong starburst with a star formation rate $\approx$\,500-1000\,M$_{\sun}$/yr. At the sensitivity of these studies, 
$\sim$1-1.5\,mJy at 1.2\,mm, 70\% of the QSOs are not detected, and for the whole sample the behavior of the FIR luminosity with respect to the bolometric luminosity 
L$_{\rm bol}$ is similar to that of all QSOs at any redshift, pointing to an important contribution of the AGN to the FIR emission. 

Besides SDSS, the Canada-France High-z Quasar Survey (CFHQS) 
is the second largest provider of $z\sim 6$ QSOs. With 20 sources it accounts for 
about one third of the total number of QSOs at $5.7<z<6.5$. 
Coming from deeper optical surveys, the CFHQS sample contains QSOs optically much fainter than the main SDSS sample and also the SDSS deep southern survey. 
The purpose of this Note is to publish the results of the 1.2\,mm survey of the  20 
$z\sim 6$ CFHQS QSOs that we performed to extend the existing studies of FIR properties of QSOs at similar reshifts 
to optically fainter sources. For four of these sources, 1.2\,mm results were already published in Willott et al. (2007). The exceptionally strong 1.2\,mm flux density that we found  for CFHQS J1429+5447,  led to the selection of  
this source to search for CO(2-1) emission (Wang et al. 2011b). Together with those of Wang et al. (2011a), our results provide a useful background for the much deeper studies of submillimeter properties of optically faint $z\sim 6$  QSOs 
that we have already begun with The Atacama Large Millimeter/submillimeter Array (ALMA)  (Willott et al.\ in prep.).

 Cosmological parameters of $H_0=70~ {\rm km~s^{-1}~Mpc^{-1}}$, $\Omega_{\mathrm M}=0.28$ and $\Omega_\Lambda=0.72$ (Komatsu et
al. 2009) are assumed throughout.

\section{Observations}

The CFHQS is an optically-selected survey for $5.8<z<6.5$ QSOs. It was carried out in regions of the sky observed as part of the Canada-France-Hawaii Telescope Legacy Survey\footnotemark, the Subaru/XMM-Newton Deep Survey and the Red-sequence Cluster Survey 2. With $z'_{\rm AB}$ band magnitude survey limits ranging from $z'_{\rm AB}=22$ to $z'_{\rm AB}=24.5$ in different regions of the sky, CFHQS QSOs are typically $10-100$ times lower luminosity than QSOs at the same redshift from the main SDSS sample (Fan et al. 2006) and many are less luminous than those from the SDSS Deep Stripe (Jiang et al. 2008; 2009). The CFHQS survey contains 20 spectroscopically-confirmed QSOs at $5.88<z<6.43$, a significant fraction of the total of $\approx 60$ QSOs known at this epoch (Willott et al. 2007; 2009; 2010a,b). Table 1 contains the positions, redshifts, magnitudes, absolute magnitudes and bolometric luminosities of the full sample. Bolometric luminosities have been determined from the absolute magnitudes at 1450\AA\ assuming a bolometric conversion factor $\zeta_{\rm 1450A}$\,=\,L$_{\rm bol}$/($\nu$L$_{\rm \nu,1450A}$) of 4.4 (Richards et al. 2006).  

\footnotetext{http://www.cfht.hawaii.edu/Science/CFHTLS}

The millimeter observations were performed within the pool observing sessions at the IRAM 30m telescope in the winters 2007 through 2010, using the 117 element version of the Max Planck Millimeter Bolometer (MAMBO) array (Kreysa et al.\ 1998) operating at an average wavelength of 1.2\,mm (250\,GHz). We used the standard on-off photometry observing mode, chopping between the target and sky by 
32$''$ in azimuth at 2 Hz every 10\,s, and nodding the telescope every 10 or 20\,s (see e.g.\ Wang et al.\ 2011a). On average, the noise of the channel used for point-source observations was about 35-40 mJy.$\sqrt{s}$ / beam. %
This allowed us to achieve rms $\lesssim$\,0.5-1.0\,mJy for 18 of the 20 sources, with about 0.5-1.5\,hr of telescope time per source. Unfortunately, poor weather conditions during the last observing runs prevented us to reach the aimed rms of $\sim$\,0.6\,mJy for 8 sources out of 20. The data were reduced with standard procedures to minimize the sky noise with the MOPSIC package developed by Zylka (1998). 

The results are shown in Table 1. Only the peculiar source CFHQS J1429+5447 is detected with a signal to noise ratio S/N\,$>$\,5. Two other sources, CFHQS J0033-0125 (Willott et al.\ 2007) and CFHQS J0102-0218, are marginally detected with S/N = 3.1 and 2.7, respectively, thanks to special deeper observations bringing the rms below 0.4\,mJy. There are 10 other sources with rms between 0.45 and 0.65\,mJy, and five between 0.75 and 1.0\,mJy, all with S/N\,$<$\,2. 

Such a 3$\sigma$ detection rate of barely 10\% is significantly smaller than that of Wang et al.\ (2011a) who report a rate of 30\%. This difference may be partially due to our lower sensitivity - average rms 0.69\,mJy instead of 0.52\,mJy for Wang et al.'s sources with m$_{1450}$\,$>$\,20.2. However, there is certainly also an effect of the larger bolometric luminosities of Wang et al.'s sample, as discussed in Sec.\ 4.

\begin{table*}
\centering
\caption{Optical data and results of 1.2\,mm observations of full CFHQS sample}
\begin{center}
{\scriptsize }
\begin{tabular}{lcclcccrc}
\multicolumn{8}{c}{}\tabularnewline
\hline
  ~~~~~~~~~~Name         &     RA            &     DEC     & Redshift$^{\dagger}$ & m$_{1450}$* & M$_{1450}$$^{\#}$ & L$_{\rm bol}$      & S$_{\rm 250GHz}$ & z'$_{\rm AB}$ \\
                                                &                     &                &                     &                  &      & ($ \rm 10^{13}\,L_{\odot}$) &          (mJy)        &     \\
\hline
 CFHQS\,J003311$-$012524 &  00:33:11.40 &  $-01$:25:24.9   &  6.13    & 21.53 &  $-24.91$   & 0.94 &  1.13$\pm$0.36  &  22.41$\pm$0.08 \\
 CFHQS\,J005006+344522   &  00:50:06.67 &      +34:45:22.6 &  6.253   & 19.84 &  $-26.65$   & 4.67 &  0.69$\pm$0.76  &  20.47$\pm$0.03 \\
 CFHQS\,J005502+014618   &  00:55:02.91 &      +01:46:18.3 &  5.983   & 21.82 &  $-24.55$   & 0.68 &  0.43$\pm$0.46  &  22.19$\pm$0.06 \\
 CFHQS\,J010250$-$021809 &  01:02:50.64 &  $-02$:18:09.9   &  5.95    & 22.02 &  $-24.34$   & 0.56 &  1.01$\pm$0.38  &  22.30$\pm$0.08 \\
 CFHQS\,J013603+022605   &  01:36:03.17 &      +02:26:05.7 &  6.21    & 22.04 &  $-24.43$   & 0.61 &  1.18$\pm$0.97  &  22.10$\pm$0.09 \\
 CFHQS\,J021013$-$045620 &  02:10:13.19 &  $-04$:56:20.9   &  6.4323  & 22.25 &  $-24.31$   & 0.70 &  1.14$\pm$0.93  &  22.67$\pm$0.05 \\
 CFHQS\,J021627$-$045534 &  02:16:27.81 &  $-04$:55:34.1   &  6.01    & 24.15 &  $-22.24$   & 0.08 &  0.37$\pm$0.57  &  24.40$\pm$0.06 \\
 CFHQS\,J022122$-$080251 &  02:21:22.71 &  $-08$:02:51.5   &  6.161   & 21.98 &  $-24.47$   & 0.63 & $-1.48\pm$1.36  &  22.63$\pm$0.05 \\
 CFHQS\,J022743$-$060530 &  02:27:43.29 &  $-06$:05:30.2   &  6.20    & 21.41 &  $-25.05$   & 1.08 & $-0.03\pm$0.53  &  22.71$\pm$0.06 \\
 CFHQS\,J031649$-$134032 &  03:16:49.87 &  $-13$:40:32.2   &  5.99    & 21.72 &  $-24.66$   & 0.75 &  2.76$\pm$1.44  &  21.72$\pm$0.08 \\
 CFHQS\,J105928$-$090620 &  10:59:28.61 &  $-09$:06:20.4   &  5.92    & 20.75 &  $-25.60$   & 1.79 &  0.03$\pm$0.82  &  20.82$\pm$0.03 \\
 CFHQS\,J142952+544717   &  14:29:52.17 &      +54:47:17.6 &  6.1831  & 20.59 &  $-25.88$   & 2.31 &  3.46$\pm$0.52  &  21.45$\pm$0.03 \\
 CFHQS\,J150941$-$174926 &  15:09:41.78 &  $-17$:49:26.8   &  6.121   & 19.63 &  $-26.80$   & 5.41 &  0.91$\pm$0.47  &  20.26$\pm$0.02 \\
 CFHQS\,J164121+375520   &  16:41:21.64 &      +37:55:20.5 &  6.047   & 21.19 &  $-25.21$   & 1.25 &  0.25$\pm$0.47  &  21.31$\pm$0.04 \\
 CFHQS\,J210054$-$171522 &  21:00:54.62 &  $-17$:15:22.5   &  6.087   & 21.37 &  $-25.05$   & 1.08 &  0.29$\pm$0.59  &  22.35$\pm$0.09 \\
 CFHQS\,J222901+145709   &  22:29:01.65 &      +14:57:09.0 &  6.152   & 21.90 &  $-24.55$   & 0.68 &  0.82$\pm$0.80  &  22.03$\pm$0.05 \\
 CFHQS\,J224237+033421   &  22:42:37.55 &      +03:34:21.6 &  5.88    & 22.09 &  $-24.25$   & 0.51 &  0.72$\pm$0.61  &  21.93$\pm$0.04 \\
 CFHQS\,J231802$-$024634 &  23:18:02.80 &  $-02$:46:34.0   &  6.05    & 21.55 &  $-24.85$   & 0.90 &  0.54$\pm$0.56  &  21.66$\pm$0.05 \\
 CFHQS\,J232908$-$030158 &  23:29:08.28 &  $-03$:01:58.8   &  6.417   & 21.53 &  $-25.02$   & 1.05 &  0.06$\pm$0.50  &  21.76$\pm$0.05 \\
 CFHQS\,J232914$-$040324 &  23:29:14.46 &  $-04$:03:24.1   &  5.90    & 21.96 &  $-24.39$   & 0.58 & $-1.45\pm$0.63  &  21.87$\pm$0.08 \\
\hline
\multicolumn{8}{c}{}
\end{tabular}
\par\end{center}
{\scriptsize \par}
\begin{flushleft}
{\small ${\dagger}$  Redshifts are from Willott et al. (2007;2009;2010a;2010b;in prep.); Wang et al. (2011b). Redshifts to 4 decimal places are from millimeter lines, those to 3 decimal places are from broad UV MgII lines and those to 2 decimal places are from Ly-$\alpha$.\\}
{\small * Apparent magnitude at rest-frame 1450\,\AA.}\\
{\small \# Absolute magnitude at rest-frame 1450\,\AA. The values given here supersede previous published values of M$_{1450}$. They  were derived by fitting a typical QSO spectrum to the observed $J$-band magnitudes.\\}
\par\end{flushleft}
\end{table*}

\section{Far-infrared luminosities}

\subsection{Estimates of $\rm  L_{FIR}$ and $M_\mathrm{dust}$}

One may convert the 250\,GHz flux densities - corresponding to $\lambda$$_{\rm rest}$ $\sim$ 170\,$\mu$m - to FIR luminosities by assuming a model for the FIR SED. For homogeneity, we take the same assumptions as Wang et al. (2008; 2011a) for this SED, i.e. assume an optically thin graybody with a dust temperature of $\rm T_d=47\,K$ and emissivity index of $\rm \beta=1.6$. As proposed by Beelen et al.\ (2006), these are typical values for the FIR  luminous QSOs at $z\sim 2$-4. We also select the same wavelength range, $\rm 42.5-122.5\,\mu m$, for the definition of the FIR luminosity as Wang et al. (2008; 2011a). Values of L$_{\rm FIR}$ thus calculated
for detected or tentatively detected sources are reported in Table 2. For the average redshift of our sample, $\langle z\rangle = 6.09$, the conversion factor between the average values of the 250\,GHz flux density and $\rm  L_{FIR}$ is 

\begin{equation}
 \langle L_{\rm FIR}\rangle/10^{12}L_{\odot} = 2.30\, \times \langle S_{\rm 250GHz}\rangle (mJy) 
\label{eq1000}
\end{equation}

For other redshifts the conversion factor is slightly different, varying
from 2.35 for $z = 5.88$ to 2.16 for $z = 6.43$.

However,  
we note that our faint sources have much lower mm fluxes
than the typical sources used to determine the dust temperature value $\rm T_d=47\,K$. If our
sources instead have parameters closer to those of nearby luminous
infrared galaxies (LIRGs, $10^{11} - 10^{12} \, {\rm L}_\odot, T_{\rm
  d}\approx 33$\,K, U et al.\ 2012),  then the values of $L_{\rm FIR}$
would be $\sim$3 times lower for $\rm T_d=33\,K$.


The dust mass $M_\mathrm{dust}$ at $T_\mathrm{dust}$ is related to the FIR luminosity by  $\rm M_{dust}=L_{FIR}/4\pi\int\kappa_{\nu}B_{\nu}d\nu$, where $\rm B_{\nu}$ is the Planck function and $\rm \kappa_{\nu}=\kappa_{0}(\nu/\nu_{0})^{\beta}$ is the dust absorption coefficient. Using 
$\rm \kappa_{0}=18.75\,cm^2g^{-1}$ at 125$\,\mu$m (Hildebrand 1983) as Wang et al. (2008; 2011a), yields   

\begin{equation}
\rm M_{dust}/10^{8}\,M_{\odot}\,=\,0.40\, \times  L_{FIR}/10^{12}\,L_{\odot}.\label{eq1000}
\end{equation}

For $\rm T_d=33\,K$, the values of $\rm M_{dust}$ would be $\sim$7 times larger than for $\rm T_d=47\,K$ for the same $L_{\rm FIR}$ ($\sim$2.4 times larger for the same $S_{\rm 250GHz}$).

\begin{table*}
\centering
\caption{Average properties}
\begin{center}
{\scriptsize }
\begin{tabular}{lcccccc}
\multicolumn{7}{c}{}\tabularnewline
\hline
 Group & Number & $\rm \left\langle L_{bol}\right\rangle$$^{*}$$^{av}$ & $\rm \left\langle L_{bol}\right\rangle$$^{*}$$^{med}$ & $\rm \left\langle S_{250GHz}\right\rangle^\dagger$
& $\rm \left\langle L_{FIR}\right\rangle^{\dagger \dagger}$ & $\rm \left\langle M_{dust}\right\rangle^{\dagger \dagger}$\\
    &     & ($\rm 10^{13}\,L_{\odot}$) & ($\rm 10^{13}\,L_{\odot}$) & (mJy) & ($\rm 10^{12}\,L_{\odot}$) &  ($\rm 10^{8}\,M_{\odot}$) \\
\noalign{\smallskip} \hline \noalign{\smallskip}
 a) all objects 								&     20  & 1.31  &  0.90    &  0.63$\pm$0.14  & 1.45$\pm$0.32 &   0.58  \\
 b) homogeneous 								&     18  & 1.38  &  0.94    &  0.64           & 1.46          &   0.59  \\
 c) regular        		    			&     19  & 1.26  &  0.75    &  0.41$\pm$0.14  & 0.94$\pm$0.32 &   0.38  \\
 d) 250\,GHz-undetected     		&     17  & 1.32  &  0.75    &  0.34$\pm$0.15  & 0.78$\pm$0.34 &   0.31  \\
 e) regular m$_{\rm 1450}<21.7$ &     9   & 2.02  &  1.08    &  0.42$\pm$0.19  & 0.96$\pm$0.44 &   0.39  \\
 f) regular m$_{\rm 1450}>21.7$ &     10  & 0.58  &  0.63    &  0.41$\pm$0.22  & 0.94$\pm$0.51 &   0.38  \\
CFHQS J1429+5447$^{\#}$         	&     1       & 2.31  &          &  3.46$\pm$0.52  & 7.85$\pm$1.18 &   3.16  \\
CFHQS J0033-0125             	&     1       & 0.94  &          &  1.13$\pm$0.36  & 2.58$\pm$0.82 &   1.04  \\
CFHQS J0102-0218             	&     1       & 0.56  &          &  1.01$\pm$0.38  & 2.35$\pm$0.89 &   0.95  \\
\hline
\multicolumn{7}{c}{}
\end{tabular}
\end{center}
\begin{flushleft}
{\small *: Bolometric luminosity: $^{av}$ mean value (straight average, with equal weight for all sources); $^{med}$ median value \\}
{\small $\dagger$: Averages of $\rm S_{250GHz}$ are performed as described in Sec.\ 3.2: groups  a to f: \\}
{\small a): Whole sample, regularized rms weigthed average.\\}
{\small b): All objects but CFHQS J0221$-$802 and CFHQS J0316$-$1340 whose 250\,GHz rms are peculiarly large; straight average (with equal weight for all sources). \\}
{\small c): All objects but CFHQS J1429+5447 which is exceptionally strong at 250\,GHz and radio loud (Sec.\ 3.2); regularized rms.  \\}
{\small d): 
All objects but CFHQS J0033-0125, CFHQS J0102-0218 and CFHQS J1429+5447 which are (tentatively) detected; regularized rms. \\}
{\small e): Same as c), but m$_{\rm 1450}<21.7$. \\}
{\small f): Same as c), but m$_{\rm 1450}>21.7$. \\}
{\small ${\#}$: The value used for $\rm L_{FIR}$ of CFHQS J1429+5447 corresponds to the totality of the measured 250\,GHz flux density, while only an unknown part is emitted by the starburst dust in the QSO host galaxy.\\}
{\small $\dagger \dagger$: $\rm \left\langle L_{FIR}\right\rangle$ is inferred from $\rm \left\langle S_{250GHz}\right\rangle$ through Eq.\ 1 (for all groups with several sources, the average redshift $\langle z \rangle$\,=\,6.09 is assumed);  $\rm \left\langle M_{dust}\right\rangle$ is inferred from $\rm \left\langle L_{FIR}\right\rangle$ through Eq.\ 2. \\}
\par\end{flushleft}
\end{table*}

\begin{figure*}
\begin{center}
{\small \includegraphics[scale=0.8]{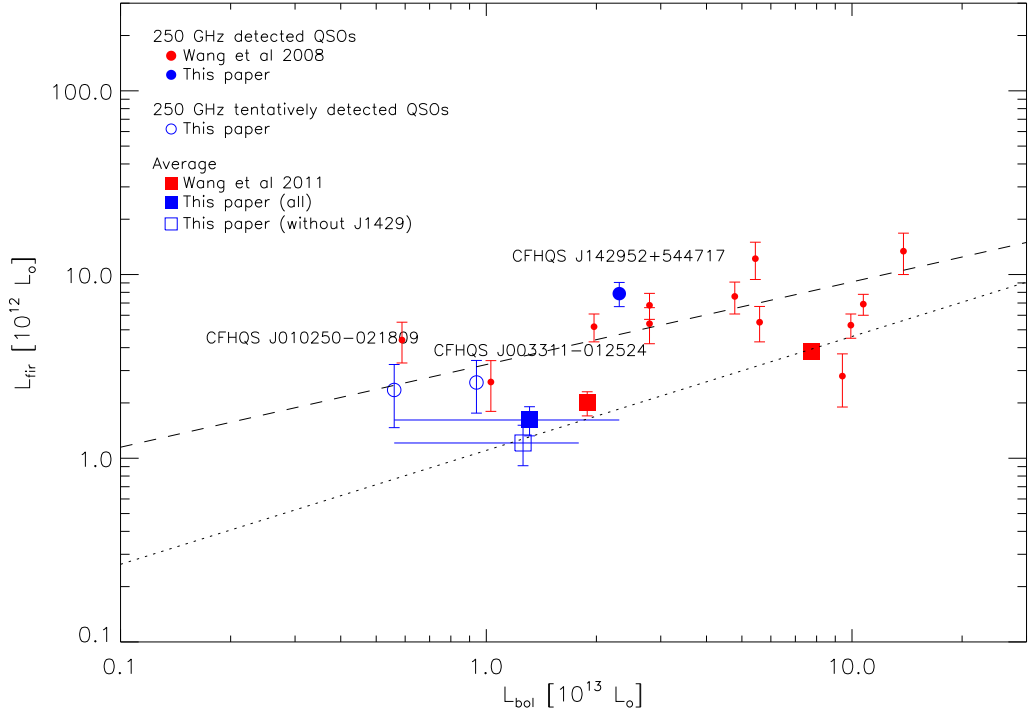}}
\caption{{\it Similar to Fig.\ 4 of Wang et al. (2011a)}. FIR and bolometric luminosity correlations of z$\sim$6 QSOs, showing: 1) average luminosities 
({\it large symbols}) for:  1a) our whole sample of 20 sources and that of 19 sources taking out CFHQS J1429+5447 (the horizontal error bars contain
80\% of the data points of these samples, excluding the two highest
and lowest values of $\rm L_{bol}$); 1b) average values of Wang et al.\ (2011a) for optically faint ($\rm m_{1450}\ge20.2$) and bright ($\rm m_{1450}<20.2$) sources; 2) 
individual luminosities ({\it small symbols}) for our 250\,GHz detections and tentative detections, together with the detections of Wang et al.\ (2008) for comparison. Note that for CFHQS J1429+5447, the value used for $\rm L_{FIR}$ corresponds to the totality of the measured 250\,GHz flux density, while only an unknown part is emitted by the starburst dust in the QSO host galaxy. The two straight lines are reproduced from Fig.\ 4 of Wang et al.\ (2011a): The dotted line [$\rm log(L_{FIR})=0.62log(L_{bol})+3.9$] is a 
power-law fit to the average luminosities  of all the QSO high-z 
samples, and the dashed line [$\rm log(L_{FIR})=0.45log(L_{bol})+6.6$] shows the power-law fit to the submillimeter or millimeter detected QSOs in all high-z samples and local ULIRGs. To be consistent with our selected bolometric conversion factor
$\zeta_{\rm 1450A}$\,=\,4.4 (Sec. 2), we have scaled all Wang et al.'s
values of $\rm L_{bol}$ using our bolometric conversion factor instead
of $\zeta_{\rm 1450A}$\,$\approx$\,6 which we inferred from Table 4 and
Fig.\ 4 of Wang et al.\ (2011a).}
\par\end{center}
\end{figure*}

\subsection{Average FIR Luminosities}

Considering the small number of 1.2\,mm detections, we use the additional information provided  
by the averages derived from various stacking of the observed 1.2\,mm flux densities.
However, some care must be taken in carrying out such averages because of the inhomogeneity of our data.
As the redshift range of our sample is small, it is simpler to perform all the averages on the 1.2\,mm flux density, $\rm S_{250GHz}$, and to infer the corresponding average of $\rm L_{FIR}$ by using the conversion factor of Eq.\ 1 for the average redshift 6.09. We may consider 
the following various averages of $\rm S_{250GHz}$:

- a) Classical rms weighted averages with weights proportional to 1/rms$^2$. However, this could give too much weight to the two sources with rms\,$<$\,0.4\,mJy whose integration time was anomalously long. It seems thus better to replace their rms by a typical value - median of the rms of the other sources (regularized rms).

- b) Plain, straight average of the  nominal values of $\rm S_{250GHz}$ with equal weights irrespectively of the rms. However, this does not take into account the difference in quality of these values. Therefore, we give in Table 2 the result of such a straight average with discarding two sources whose 250\,GHz rms are peculiarly large.

- c) As there is clearly one exceptional source, CFHQS J1429+5447, which is more than three times stronger at 1.2\,mm than all the others, and is radio loud (Sec. 3.3), one may prefer to exclude it, stacking only the 19 other sources with weights as in a).

- d) One might also exclude the two other tentative detections, stacking only 17 sources with weights as in a).

- e) \& f) One may finally try to split the sample into two halves with respect to the UV luminosity to see whether the latter can have some visible effect on $\langle$$\rm S_{250GHz}$$\rangle$.

In Table 2, we give the results corresponding to all these options. It is seen that the results of the first two do not significantly differ. As expected, dropping CFHQS J1429+5447  leads to a decrease of the average flux density, 
yielding  $\langle$S$_{\rm 250GHz}$$\rangle$\,=\,0.41$\pm$0.14 for the rms weighted average of the 19 other sources. This is probably the best average for the most representative sources of our sample.  But at 2.9$\sigma$ it is hardly meaningful. In addition small residual systematic errors in MAMBO results are not excluded at this level.  

Note that the average of $\rm log(L_{FIR})$ (or $\rm S_{250GHz}$) for groups e) and f) is found to be about the same 
despite the fact these groups are split by luminosity. This may appear surprising because one may 
expect to see a decrease of $\rm \left\langle L_{FIR}\right\rangle$ with $\rm \left\langle L_{bol}\right\rangle$ in line with the results of  Wang et al. (2011a) (see lower dotted line in Fig. 1). However this can be easily explained by the fact we do not have a huge luminosity
range in our sample. The average $\rm \left\langle L_{bol}\right\rangle$ of these groups differ by a factor of 4 and
the medians less than a factor of 2. Given each stack is detected at only S/N\,$\sim$\,2, our results are consistent with the relation of Wang et al.

\subsection{CFHQS J142952.17+544717.6}

This QSO is the only one detected at 1.2\,mm with a high S/N ratio, $\rm S_{250GHz}=3.46\pm 0.52\,mJy$. The corresponding values of the  FIR luminosity and the dust mass are reported in Table 2. Besides being one of the mm-brightest z$\sim$6 QSOs and one of the UV most luminous of our sample, it is remarkable in many respects. It is one of the only four known radio-loud $z\sim6$ QSOs (see e.g. Frey et al.\ 2011) and 
has the highest redshift and strongest radio emission among them. The EVLA observation of its CO(2-1) line (Wang et al.\ 2011b), prompted by our 1.2\,mm result, has shown a strong CO emission resolved in two peaks, one on top of the QSO position and the other 1.2$''$ away. Each corresponds to a gas mass larger than $\rm 10^{10}\,M_{\odot}$. The value of the ratio of the FIR to CO luminosities, $\rm L_{FIR}$/$\rm L'_{CO}$\,=\,300, is just below the average of this ratio for the $z\sim6$ QSOs where Wang et al. (2010) have detected CO. This indicates that probably most of the 1.2\,mm continuum emission comes from dust and not from the radio source. The latter is compact and has a very steep radio spectrum (3.03$\pm$0.005, 0.99$\pm$0.006 and 0.257$\pm$0.015 mJy at 1.6, 5 and 32 GHz, respectively; Frey et al.\ 2011; Wang et al.\ 2011b). This is another indication that most of the millimeter emission is due to dust, 
although a significant synchrotron contribution cannot be fully excluded.
Moreover, it is not yet known how the observed 250\,GHz flux density is shared between the dust emission of each galaxy, and possibly the synchrotron emission of the QSO. Therefore in Table 2 and Fig.\ 1, the value of $\rm L_{FIR}$ corresponds to the totality of the measured 250\,GHz flux density, while only an unknown part is emitted by the starburst dust in the QSO host galaxy.

\section{Discussion and conclusions}

The most natural explanation for an excess of FIR emission is star formation. If the FIR emission is powered by a dusty starburst, there is a direct relation between 
SFR and $\rm L_{FIR}$. The most generally used relation is given by Kennicutt (1998), SFR($\rm M_{\odot}$/yr)\,=\,1.72 $\times$ 10$^{-10}$ $\rm L_{FIR}( L_{\odot}$) for a Salpeter IMF and 8\,$< \lambda < 1000\,\mu$m. However, the exact conversion factor significantly depends on the IMF and the selected wavelength range for estimating $\rm L_{FIR}$. With our selected wavelength range, 42.5-122.5\,$\mu$m, it is conservative to assume that SFR($\rm M_{\odot}$/yr)\,$\gtrsim$\, 1.5 $\times$ 10$^{-10}$ $\rm L_{FIR}( L_{\odot}$).

For sources with large FIR excess, such as CFHQS J1429+5447, it is generally agreed that most of the FIR emission is powered by a starburst (see e.g.\ Beelen et al.\ 2006; Wang et al.\ 2008; Mor et al.\ 2012). This yields SFR\,$\gtrsim$\,500\,$\rm M_{\odot}$/yr for each of the two galaxies of CFHQS J1429+5447. 
 For the other sources, making the assumption that all the FIR emission is due to star formation, the value of Table 2 (group c), $\langle$L$_{\rm FIR}\rangle$\,$\approx$\,0.94$\pm$0.32 $\times$ 10$^{12}$\,L$_{\sun}$, would imply SFR\,$\gtrsim$\,140$\pm$50\,$\rm M_{\odot}$/yr.

For the much smaller FIR emission of the majority of QSOs, an important starburst contribution is also likely; however, it is possible that a fraction also comes from dust powered by the AGN. This was pointed out for z\,$\sim$\,6 QSOs by Wang et al. (2011a). They extended their discussion to 
lower redshifts as well as various authors that they quote (see also e.g.\ Rosario et al.\ 2012; Dai et al.\ 2012, for more recent references). In their Fig.\ 4,  (partially reproduced in Fig.\ 1), they have shown that there is a remarkably uniform correlation of $\rm L_{FIR}$ with $\rm L_{bol}$ for QSO average luminosities, both for various samples of  high-$z$ QSOs, and of low-$z$ ones (e.g. Hao et al.\ 2005). This correlation is well represented by the  power law [$\rm log(L_{FIR})=0.62\times log(L_{bol})+3.9$] (see Fig.\ 1). A starburst contribution is needed to explain the FIR slope of 0.6, while the slopes for IRAS 12\,$\mu$m  and 25\,$\mu$m of local QSOs are linear and consistent with AGN heating (Hao et al.\ 2005). However, such a relatively high value of 0.6 for the FIR slope may be evidence for a combination of AGN and starburst contributions.

In Fig.\ 1, similar to Fig.\ 4 of Wang et al. (2011a), we show the correlations for our sample between $\rm L_{FIR}$ and 
$\rm L_{bol}$, computed as described above (Sec.\ 2). Comparing the plotted average luminosities for all our 20 sources, with and without CFHQS J1429+5447, to the averages of Wang et al. (2011a) for m$_{1450}\,<\,20.2$ and $>\,20.2$, shows that our sample represents a significant extension to the work of Wang et al.\ toward optically weaker sources. Our average for all 20 observed sources, including CFHQS J1429+5447, appears consistent with the two averages of Wang et al.\ for all the $z\sim6$ QSOs they observed, respectively optically faint and bright. The average for our 19 sources, without CFHQS J1429+5447, is in better agreement with the power-law fit that Wang et al.\ found for the average luminosities  of all the high-$z$ samples of QSOs observed in sub/millimeter.

The position of CFHQS J1429+5447 in Fig. 1 is well within the region of 1.2\,mm detected sources by Wang et al.\ (2008; 2011a).
Finally in Fig.\ 1 we have added the positions of the two tentatively detected sources at 1.2\,mm, CFHQS J0033$-$0125 (S/N\,=\,3.1) and CFHQS J010250-0218 (S/N\,=\,2.7). They are again close to the line fitting mm-detected sources.

In conclusion, our 1.2\,mm observations confirm the results of Wang et al. (2011) about the FIR emision of $z\sim 6$ QSOs and extend them toward optically fainter sources. The core of our sample is made of such faint sources, 21.5\,$\lesssim$~z'$_{\rm AB}$~$\lesssim$\,22.7, corresponding to $\rm L_{\rm bol}$\,$\sim$ (0.5-1) $\times$ 10$^{13}$\,L$_{\sun}$ and black-hole masses of a few 10$^8$\,M$_{\sun}$ (Willott et al.\ 2010b). For such QSOs, the average FIR luminosity is weak, but probably still significant $\langle$L$_{\rm FIR}\rangle$\,$\approx$\,0.94$\pm$0.32 $\times$ 10$^{12}$\,L$_{\sun}$. This corresponds to an average star formation rate of a few 100's M$_{\sun}$/yr. However, there is certainly a large dispersion for individual sources around these average values - at least a factor $\sim$3 in both directions, as examplified by our two tentative 1.2\,mm detections (CFHQS J0033$-$0125 and CFHQS J0102-0218) in this L$_{\rm bol}$ range, and the very low 1.2\,mm flux density  measured with ALMA for two CFHQS QSOs (Willott et al. in prep.).

Such low star formation rates are probably reflecting small gas and total masses of the host galaxy. This should 
then favor large ratios ${\rm M_{BH}/M_{galaxy}}$, which could be significantly larger than the typical ratio at $z=0$ (${\rm M_{BH}/M_{bulge}\,\approx 0.0014}$; Marconi \& Hunt 2003). But such ratios remain very uncertain in the absence of direct measurements of  ${\rm M_{gas}}$ and ${\rm M_{galaxy}}$, e.g. by CO or C$^+$ observations (e.g.\ Wang et al.\ 2011b; 2012; Willott et al.\ in prep.).

\begin{acknowledgements}
We thank T.\ Forveille 
for useful discussions. Based on observations with the IRAM 30m MRT at
Pico Veleta. IRAM is supported by INSU/CNRS (France),
MPG(Germany) and IGN(Spain). Thanks to the queue observers
at IRAM who obtained data
for this paper. Thanks to INSU for supporting a visit to France of G. Orellana. Based on observations obtained with MegaPrime/MegaCam, a joint project
of CFHT and CEA/DAPNIA, at the Canada-France-Hawaii Telescope (CFHT)
which is operated by the National Research Council (NRC) of Canada,
the Institut National des Sciences de l'Univers of the Centre National
de la Recherche Scientifique (CNRS) of France, and the University of
Hawaii. This work is based in part on data products produced at
TERAPIX and the Canadian Astronomy Data Centre as part of the
Canada-France-Hawaii Telescope Legacy Survey, a collaborative project
of NRC and CNRS. 
\end{acknowledgements}

\end{document}